\documentclass[ onecolumn, prd, showpacs]{revtex4}

\usepackage{bm}
\usepackage{graphicx}
\usepackage{psfrag}

\newcommand{\dd}{\ensuremath{\textrm{d}}}
\newcommand{\Ee}{\ensuremath{\textbf{E}^3}}

\newcommand{\UD}[2]{\ensuremath{^{#1}_{\phantom{#1} #2}}}

\newcommand{\pard}[2]{\ensuremath{\frac{\partial #1}{\partial #2}}}

\newcommand{\PP}{\ensuremath{{\cal P}}}
\newcommand{\parital}{\partial}

\newcommand{\Dd}{\ensuremath{ {\cal D} }}
\newcommand{\Nn}[1]{\ensuremath{ {\cal N}\left[ {#1} \right] }}

\newcommand{\Pp}{\ensuremath{ {\cal P} }}

\newcommand{\Npm}[1]{\ensuremath{ {\cal NR}\left[ {#1} \right] }}

\newcommand{\av}[1]{\ensuremath{\langle{#1}\rangle}}

\newcommand{\leftsym}[1]{\ensuremath{\left({#1}\right.}}
\newcommand{\rightsym}[1]{\ensuremath{\left.{#1}\right)}}
\newcommand{\leftasym}[1]{\ensuremath{\left[ {#1} \right.}}
\newcommand{\rightasym}[1]{\ensuremath{\left. {#1} \right] }}

\newtheorem{theorem}{Theorem}[section]

\begin{document}

\title[Coarse--graining of inhomogeneous...]{Coarse--graining of inhomogeneous dust flow in General Relativity via isometric embeddings}


\author{Miko\l{}aj Korzy\'nski}
\address{Max--Planck-Institut f\"ur Gravitationsphysik\\ (Albert--Einstein--Institut) \\ 14476 Golm, Germany}

\begin{abstract}

We present a new approach
to coarse-graining of variables describing dust flow in GR.
It is based on assigning quasi-local shear, twist and expansion to
2-dimensional surfaces with the help of isometric embeddings into the 3--dimensional Euclidean space and deriving the time evolution equations for them. In contrast to the popular Buchert's scheme it allows to
coarse--grain tensorial quantities in a coordinate--independent way. The framework can be used to
estimate backreaction in inhomogeneous cosmological models. 

\end{abstract}

\maketitle


\section{Introduction}

In this paper we discuss an alternative to the well--known Buchert's averaging scheme \cite{Buchert:1999mc} for inhomogeneous cosmological models with dust. The new procedure allows for coarse--graining of both scalar (expansion) and tensorial (shear and vorticity) part of the fluid velocity gradient in a coordinate--independent way. Given a 3+1 decomposition of the spacetime we assign the coarse--grained expansion, shear and vorticity to three--dimensional comoving domains of the fluid. The resulting non--local quantities depend on the 3+1 splitting, but not on the coordinate system introduced on the spatial slices. 
If the fluid is irrotational, there exists a preferred, orthogonal splitting and the formalism is entirely covariant; otherwise one needs to fix the constant time slices by an additional condition.

The procedure is based on a reformulation of the definition of the volume average of the velocity gradient over a comoving domain in Newtonian theory. It turns out that it is possible to reexpress that volume average as a functional of the time derivative of the metric tensor induced on the domain's two--dimensional boundary. It has been noted before \cite{buchert-1997-320, ehlers-1997-29} that such volume--averaged velocity gradients satisfy an evolution equation which is very similar to the evolution equation of the local velocity gradient, the only difference being a surface integral of the inhomogeneities of the flow over the domain's boundary. This volume average can be rewritten as a \emph{surface} integral of the fluid velocity due to the divergence theorem. The key observation is that the surface values of the velocity field can be almost uniquely reconstructed  from the comoving time derivative of the metric induced on the domain's boundary as long as the that boundary has the topology of $S^2$ and its metric has a positive curvature.  The only part of the velocity field which cannot be reconstructed is the total linear and angular velocity, responsible for translations and rotations of the surface as a whole. However, since they do not contribute to the average shear and expansion, these quantities can be evaluated just from the reconstructed velocity field. Thus it is possible to reexpress in Newtonian theory the average shear and expansion as functionals of the derivative of the induced metric on the boundary.

The reformulated definition carries over easily to the relativistic case. If we fix a comoving, three--dimensional domain in a 3+1 decomposed spacetime, we can find the isometric embeddings of the domain's two--dimensional boundaries into the three--dimensional Euclidean space and evaluate the coarse--grained expansion and shear. If the background metric is not flat, the resulting quantities are not equivalent to any volume integrals any more, but nevertheless they can serve as a measure the deformation of a finite fluid element. As we show in the paper, the new definition satisfies simple reasonability conditions: it yields the right answer in the limit of the domain shrinking to a point and in FLRW metrics. 

We also present a way to define the coarse--grained  vorticity from the projection of particle four--velocities to
the spatial constant--time slice if the dust is rotating. The definition is independent of the coordinates introduced on the spatial slice, but in contrast to the irrotational case we do not have a simple, unique definition of a preferred 3+1 splitting and the coarse--grained quantities do depend on it.

Finally we present the evolution equation for the coarse--grained velocity gradient and show that its structure is very similar to the evolution equation for its local counterpart. Two additional terms in the form of surface integrals may serve as a measure of backreaction. i.e. the impact of the inhomogeneities of the flow on the large scale motion of the dust within the domain.

The proposed procedure is more mathematically involved then Buchert's, as it requires finding the isometric embeddings of $S^2$ surfaces, which is equivalent   to   solving a system of coupled, non--linear PDE's. In most cases this is impossible to do analytically, but numerical methods have already been developed for that problem \cite{bondarescu-2002-19, nollert}. Let us also note that in contrast to Buchert's approach we do not have evolution equations for the coarse--grained curvature and matter density. The dust was chosen as the matter model in this paper just for its simplicity; the formalism introduced here can be applied to any other fluid in General Relativity with only minor modifications. For a more exhaustive treatment of the topic, including rigorous derivations of the results, see \cite{korzynski--2009}.

\section{Coarse--graining via isometric embeddings}

Consider gravitating, Newtonian, presureless dust in a flat space. It is described by the velocity field $v^i$, density $\rho$ and the Newtonian potential $\phi$. The gradient of the velocity field $Q\UD{i}{j} = v\UD{i}{,j}$ 
 describes how an infinitesimal fluid or dust element is deformed and how it rotates during its motion. It can be decomposed into
the scalar expansion, symmetric traceless shear and antisymmetric vorticity 
\begin{eqnarray*}
 Q_{ij} = \frac{1}{3}\,\theta\,\delta_{ij} + \sigma_{(ij)} + \omega_{[ij]}.
\end{eqnarray*}
$Q_{ij}$ satisfies the evolution equation along the flow
\begin{eqnarray}
 \frac{D}{\partial t}\,Q\UD{i}{j} = -Q\UD{i}{k}\,Q\UD{k}{j} - \phi\UD{,i}{,j},\label{eqlocalQevol}
\end{eqnarray}
where $\frac{D}{\partial t}$ is the convective derivative (see \cite{korzynski--2009, buchert-1997-320, ehlers-1997-29}). This equation has a coarse--grained counterpart
which one obtains by taking the volume average of $Q\UD{i}{j}$ over a comoving domain $G_t$
\begin{eqnarray}
 \av{Q\UD{i}{j}} = \frac{1}{V}\,\int_{G_t} Q\UD{i}{j}\,\dd^3x = \frac{1}{V}\,\int_{G_t} v\UD{i}{,j}\,\dd^3 x \label{eqQ}. 
\end{eqnarray}
Note that $\av{Q\UD{i}{j}}$ is effectively a surface integral over the boundary of $\partial G_t$
\begin{eqnarray}
 \av{Q\UD{i}{j}} = \frac{1}{V}\,\int_{\partial G_t} v^i\,n_j\,\dd\sigma \label{eqQsurface}.
\end{eqnarray}
Its evolution equation has a similar structure to (\ref{eqlocalQevol}),
\begin{eqnarray}
 \pard{}{t}\,\av{Q\UD{i}{j}} = -\av{Q\UD{i}{k}}\,\av{Q\UD{k}{j}} - \av{\phi\UD{,i}{,j}} + B\UD{i}{j} \label{eqavQevol}
\end{eqnarray}
with $\av{\phi\UD{,i}{,j}}$ being the volume average of the Hessian of the potential over $G_t$.
The additional term
\begin{eqnarray}
 B\UD{i}{j} = \frac{1}{V}\,\int_{\partial G_t} (\delta v^k\,\delta v\UD{i}{,j}\,n_k - \delta v^k\,\delta v\UD{i}{,k} \,n_j)\,\dd\sigma, \label{eqnewtonbackr}
\end{eqnarray}
measures how the inhomogeneities of the fluid velocity field $\delta v^i = v^i - \av{Q\UD{i}{j}}\,x^j$ affect its
large scale motion in the domain $G_t$ \cite{buchert-1997-320, ehlers-1997-29,korzynski--2009}.

In the relativistic case, on a non--flat background, we do not have a well--defined, coordinate--independent notion of volume averages of tensor objects, so we cannot use (\ref{eqQ}) to obtain a relativistic counterpart of (\ref{eqavQevol}).
It turns out, however, that definition (\ref{eqQ}) can be reformulated in a geometric, coordinate independent manner which has a  natural extension to non--flat backgrounds.
We will first deal with the symmetric part of $\av{Q_{ij}}$, i.e. shear and expansion. We will show that under certain technical assumptions they can be reexpressed as functionals of the metric induced on the boundary $\partial G_t$ and its convective (comoving with the dust) time derivative.

The main tool we are going to employ is the isometric embedding theorem for surfaces of $S^2$ topology, conjectured by Weyl and proved by Lewy, Alexandrov, Pogorelov, Nirenberg and Cohn--Vossen (see \cite{spivak-chapters}, \cite{han-hong}):
\begin{theorem}[Isometric embedding theorem for $S^2$]
 Given a compact, orientable surface $S$ homeomorphic to $S^2$, with positive metric $q$ whose scalar
curvature $R > 0$. Then 
\begin{itemize}
 \item there exists an isometric embedding
\begin{eqnarray*}
 f: S \mapsto \Ee 
\end{eqnarray*}
into the 3--dimensional Euclidean space;
\item the embedding is unique up to rigid rotations, translations and reflexions.
\end{itemize}\label{thiso}
\end{theorem}
If we fix the orientation, the theorem states that for a given $S^2$ surface with a positive metric of positive curvature can be recognized as a submanifold in $\Ee$ and this can be done only in one way, up to moving the surface around and rotating it as a whole. Surfaces satisfying the hypothesis of theorem \ref{thiso} will be
called admissible.

Consider now the boundary $\partial G_t$ of the coarse--graining domain, parametrized by two coordinates $\theta^A$ labeling individual particles. It is described, 
together with its time evolution, by equations
\begin{eqnarray*}
 x^i = \zeta^i(t,\theta^A). 
\end{eqnarray*}
The induced metric has the form of
\begin{eqnarray}
 q_{AB}(t,\theta^A) = \zeta\UD{i}{,A}\,\zeta\UD{j}{,B}\,\delta_{ij} \label{eqmetric1}
\end{eqnarray}
and its time derivative in coordinates $\theta^A$
\begin{eqnarray*}
 \dot q_{AB} = 2v\UD{i}{\leftsym{,A}}\,\zeta\UD{j}{\rightsym{,B}}\,\delta_{ij}.
\end{eqnarray*}

Assume now that $\partial G_t$ is admissible at least for some time and that we know its position in space at one instance of one time $t=t_0$. We will show that in that case $\av{Q_{(ij)}}$ can be read out solely from the induced metric (\ref{eqmetric1}), even if we do not have each individual particle's velocity and position.

Given $q_{AB}(t)$ we can find the isometric embeddings of $\partial G_t$, consistent with its orientation, into \Ee at each time $t$. We assume that at $t_0$ the embedding is known, but at other times we need to solve the following system of PDE's
 \begin{eqnarray}
 q_{AB} = \chi\UD{i}{,A}\,\chi\UD{j}{,B}\,\delta_{ij} \label{eqmetric}
\end{eqnarray}
for three functions $\chi^i(t,\theta^A)$.
 We
construct this way a one--parameter family of embeddings 
\begin{eqnarray*}
 x^i = \chi^i(t,\theta^A)
\end{eqnarray*}
for which we assume the additional condition 
\begin{eqnarray*}
 \chi^i(t_0,\theta^A) = \zeta^i(t_0,\theta^A).
\end{eqnarray*}
The first statement of theorem \ref{thiso} assures that functions $\chi^i(t,\theta^A)$ can always be found, although they are non--unique.
The true particle positions of $\partial G_t$, given by $\zeta^i(t,\theta^A)$, constitute trivially a family of isometric embeddings, so from the 
the uniqueness part of \ref{thiso} we know that functions $\chi^i$  and $\zeta^i$ are related  via
\begin{eqnarray*}
 \chi^i(t,\theta^A) = R\UD{i}{j}(t)\,\zeta^j(t,\theta^A) + W^i(t),
\end{eqnarray*}
 where $R\UD{i}{j}$ is a rotation. If we take a time derivative of this equation at $t=t_0$, we obtain
\begin{eqnarray*}
 \left.\pard{\chi^i}{t}\right|_{t=t_0} = u^i(\theta^A) = v^i(t_0,\theta^A) + \Omega\UD{i}{j}\,\zeta^j(t_0,\theta^A) + E^i.
\end{eqnarray*}
with constants $\Omega_{ij} = -\Omega_{ji}$ and $E^i$.
Thus the velocity field $u^i$, reconstructed from an arbitrary family isometric embeddings, differs from the true one only by the restrictions to
$\partial G_t$ of a rotational ($\Omega\UD{i}{j}\,x^j$) and s translational ($E^i$) vector field. 
This difference does not influence the value of integral (\ref{eqQsurface}) symmetrized in the two indices
\begin{eqnarray}
 \int_{\partial G_t} u_{\leftsym{i}} n_{\rightsym{j}}\,\dd \sigma = \int_{\partial G_t} v_{\leftsym{i}} n_{\rightsym{j}}\,\dd \sigma = \av{Q_{(ij)}}, \label{eqinv}
\end{eqnarray}
so we can use the reconstructed $u$ instead of $v$ to evaluate the shear and expansion.

In fact we do not have to find the embeddings explicitly to obtain the reconstructed velocity field.  
By differentiating (\ref{eqmetric}) with respect to $t$ we obtain an equation relating  $u^i$ directly to the convective time derivative of $q_{AB}$ at $t=t_0$
\begin{eqnarray*}
 2u\UD{i}{\leftsym{,A}}\,\zeta\UD{j}{\rightsym{,B}}\,\delta_{ij} = \dot q_{AB}.
\end{eqnarray*}
The linear operator $\PP_\zeta[Y]_{AB} = 2Y\UD{i}{\leftsym{,A}}\,\zeta\UD{j}{\rightsym{,B}}\,\delta_{ij}$ appearing on the left hand side
yields the variation of the surfaces' induced metric if the embedding itself is being dragged by a vector field $Y^i$.
By the virtue of theorem \ref{thiso} for an admissible surface it has a six--dimensional kernel consisting of rigid motions and a ''non--unique inverse'' $\PP_{\zeta}^{-1}$. 
We can now rewrite the definition of the symmetric part coarse--grained velocity gradient explicitly as a functional of $\dot q_{AB}$ 
\begin{eqnarray}
 \av{Q_{(ij)}} = \frac{1}{V}\,\int_{\partial G_t} \PP_\zeta^{-1}\left[\dot q_{AB}\right]_{\leftsym{i}}\,n_{\rightsym{j}}\,\dd \sigma. \label{eqQfromdotq}
\end{eqnarray}

\section{Velocity gradient in General Relativity}
Formula (\ref{eqQfromdotq}) is clearly more complicated that (\ref{eqQ}), but its main advantage lies in the fact that it can be generalized to fluids on a non--flat background in a simple and natural way. Consider a (possibly curved) spacetime, filled with dust described by density $\rho$ and four--velocity field $u$. The integral curves of $u$ are geodesics, i.e. $\nabla_u u^\mu = 0$. The four--velocity gradient $Z_{\mu\nu} = \nabla_\nu u_\mu$ describes the deformations of an infinitesimal fluid element in its motion. It is orthogonal to $u$ in both indices and therefore effectively a three--dimensional object $Z_{\mu\nu}$ satisfies
the relativistic counterpart of (\ref{eqlocalQevol})
\begin{eqnarray}
 \nabla_u Z\UD{\mu}{\nu} = -Z\UD{\mu}{\rho}\,Z\UD{\rho}{\nu} - R\UD{\mu}{\alpha\nu\beta}\,u^\alpha\,u^\beta \label{eqdotZ}.
\end{eqnarray}
 Unless stated otherwise, in this paper we will assume the flow to be irrotational, i.e. $Z_{\mu\nu} = Z_{(\mu\nu)}$.
 We will now show how to derive the coarse--grained counterpart of (\ref{eqdotZ}) in which $\av{Z_{ij}}$ corresponds to a finite fluid volume rather than an infinitesimal element. 

\begin{figure}
  \includegraphics[height=0.2\textheight]{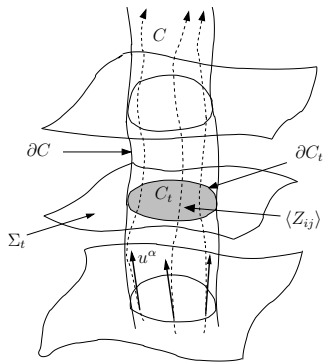}
  \caption{Coarse--graining of $Z_{\mu\nu}$ over the slices of a tube of particles}
  \label{figtube}
\end{figure}

In comoving and orthogonal coordinates $(t,y^i)$
one can perform the standard ADM decomposition of the metric into the spatial part $h_{ij}$, with vanishing $N^i$ and $N=1$ due to the coordinate system choice.
Consider a three--dimensional tube of particle worldlines $\partial C$ enclosing a cylinder $C$ (see fig. \ref{figtube}). The tube is described by equations $y^i = \xi^i(\theta^A)$, $\xi^i$ being three time--independent functions and $\theta^A$ comoving coordinates on the tube labeling individual particles. The intersections of the cylinder with constant time slices $\Sigma_t$, denoted in this paper by $\partial C_t$, constitute two--dimensional surfaces of spherical topology. Assuming that the surfaces are admissible we assign the coarse--grained velocity gradient to their interior in the following way: we take the induced metric 
\begin{eqnarray}
q_{AB}(t,\theta^A) = \xi\UD{i}{,A}\,\xi\UD{j}{,B}\,h_{ij}(t) \label{eqqab}
\end{eqnarray}
 on $\partial C_t$ and find an arbitrary family of isometric embeddings  $f_t: \partial C_t \to \parital D_t \subset \Ee$. They are described by equations
\begin{eqnarray*}
 x^a = \chi^a(t,\theta^A)
\end{eqnarray*}
with three functions $\chi^a$\footnote{Indices $a,b,c,\dots$ run from 1 to 3, just like $i,j,k,\dots$. We will reserve the former for geometric objects in \Ee and the latter for objects on a constant time slice of the spacetime.} satisfying a system of non--linear PDE's
\begin{eqnarray*}
 q_{AB}(t,\theta^A) = \chi\UD{a}{,A}\,\chi\UD{b}{,B}\,\delta_{ab}.
\end{eqnarray*}
These are subject to gauge transformations of the form of rigid motions
\begin{eqnarray}
 \chi^a(t,\theta^A) \to R\UD{a}{b}(t)\,\chi^b(t,\theta^A) + W^a(t) \label{eqgauge}.
\end{eqnarray}
We can now define a fictious rather then reconstructed velocity field $v$ on the surface image in $\Ee$ by taking the time derivative of each particle's positions given by $\chi^a$.
The vector field can also be recovered directly from $\dot q_{AB} = \xi\UD{i}{\leftsym{,A}}\,\xi\UD{j}{\rightsym{,B}}\,\dot h_{ij}(t)$ by the inverse of $\Pp_\chi$.
We can then use $v$ to evaluate the coarse--grained $Z_{(ij)}$ as tensors in $\Ee$
\begin{eqnarray}
 \av{\tilde Z_{(ab)}} = \frac{1}{V_0}\int_{\partial D_t} v_{\leftsym{a}}\,n_{\rightsym{b}}\,\dd \sigma \label{eqZab}
\end{eqnarray}
($V_0$ denotes here the volume enclosed by $\partial D_t$ in $\Ee$). Because of (\ref{eqinv}) these integrals are independent of the vector field $v$ chosen, but they do depend on the orientation of the embedded surface and under gauge transformations (\ref{eqgauge}) they transform according to
\begin{eqnarray*}
 \av{\tilde Z_{ab}} \to R\UD{c}{a}\,R\UD{d}{b}\,\av{\tilde Z_{cd}}. 
\end{eqnarray*}
\begin{figure}
  \includegraphics[height=0.2\textheight]{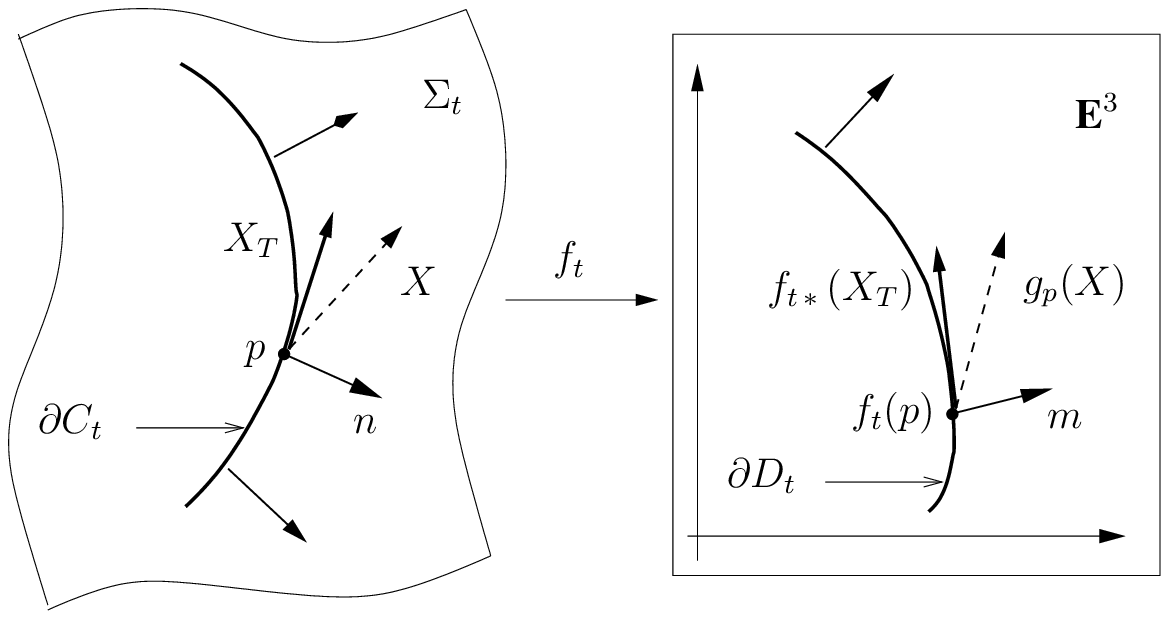}
  \caption{Canonical isometry $g_p: T_p\Sigma_t \to \Ee$}
  \label{figemb}
\end{figure}

 In order to get rid of that gauge dependence we need to map $\av{\tilde Z_{ab}}$ back to the spacetime.
Each isometric embedding $f_t$ defines canonical isometries $g_p$ between the tangent spaces $T_p \Sigma_t$ to the constant time slices at points $p \in \partial C_t$
and $\Ee$. Any vector $X \in T_p\Sigma_t$ can be decomposed into the tangent part to $\partial C_t$ and the projection to the outward pointing normal $n^i$
\begin{eqnarray*}
 X^i = X_T^i + X_n\,n^i.
\end{eqnarray*}
We then take 
\begin{eqnarray*}
  g_p(X)^a = \left(f_{t\,*}\,X_T\right)^a + X_n\,m^a,
\end{eqnarray*}
where $m^a$ is the outward pointing normal to the image of the surface at $f(p)$ (see fig. \ref{figemb}).
$g_p$ preserves the metric tensor and can be used to map geometric objects from  $T_p\Sigma_t$ to $\Ee$ or back. 
In particular, one can pull $\av{\tilde Z_{ab}}$ back to each point $p$ in $\partial C_t$ by $g_p^*$
\begin{eqnarray*}
 \av{Z_{ij}}_p = g_p^*\left(\av{\tilde Z}\right)_{ij}.
\end{eqnarray*}
It is straightforward to verify that the tensor field $\av{Z_{ij}}$ constructed on $\partial C_t$ this way is insensitive to transformations (\ref{eqgauge}) and thus completely independent of the choice of isometric embeddings $f_t$.

If the antisymmetric part of $Z_{\mu\nu}$, i.e. the vorticity, does not vanish, it can be coarse--grained as well, although in a different way. Note that the 3+1 splitting cannot be orthogonal in that case and the four--velocity field $u^\mu$ has a non--vanishing projection to the constant time slice, denoted by $w^i$. We define now the antisymmetric part of 
$\av{\tilde Z_{ab}}$ as
\begin{eqnarray}
 \av{\tilde Z_{[ab]}} = \frac{1}{V_0}\,\int_{\partial D_t} g(w)_{\leftasym{a}}\,n_{\rightasym{b}}\,\dd\sigma \label{eqQasym}.
\end{eqnarray}
Note however that, in contrast to the rotating case, in a spacetime filled with rotational dust it is difficult to single out a preferred 3+1 splitting. Unfortunately, just like in the Buchert's formalism, the values of $\av{Z_{ij}}$ depend on the splitting one chooses in general.
Throughout the rest of the paper we will only deal with the non--rotating dust and in that case $\av{Z_{[ij]}} = 0$. 

\subsection{The coarse--graining procedure in the narrow tube limit and in FLRW metrics}
Outside the Newtonian setting we loose the equivalence of $\av{Z_{ij}}$ to the volume average
of the local velocity gradient. Nevertheless the newly defined quantities turn out to have the correct limiting behavior, namely if we shrink the coarse--graining domain to a single point, we recover the local $Z_{ij}$ at that point. 
More precisely, consider a one--parameter family of tubes described by equation
\begin{eqnarray}
y^i = \lambda\,\xi^i(\theta^A). \label{eqlambda}
\end{eqnarray} 
with a positive parameter $\lambda$. As $\lambda \to 0$, the tube shrinks down to the single worldline $y^i = 0$.
One can prove that for small $\lambda$ the values of $\av{Z_{ij}}$, including the antisymmetric part if the vorticity doesn't vanish,  can be expanded at every point $p$ of the tube
in powers of $\lambda$ as 
\begin{eqnarray}
 \av{Z_{ij}} = \left.Z_{ij}\right|_{y^i=0} + O(\lambda).
\end{eqnarray}
Intuitively one can understand this fact in the following way: expressions (\ref{eqZab}) and (\ref{eqQ}) are perfectly equivalent if the background metric is flat and if the velocity gradient is constant over the coarse--graining domain. In general none of these conditions is satisfied on $\Sigma_t$, but the more the domain under consideration shrinks, the less it ``feels'' the background curvature and the inhomogeneities in $Z_{ij}$. As it shrinks down to a point, both curvature and inhomogeneities drop out and $\av{Z_{ij}}$ is equal to the volume average of the local $Z_{ij}$ over a small domain around the point $y^i=0$. It is now straightforward to prove that such volume average approaches then the
value of $Z_{ij}$ at $y^i = 0$. For a rigorous argument see \cite{korzynski--2009}.

In order to test the coarse--graining procedure we may also apply it to known homogeneous cosmological solutions. For a dust--filled Friedman--Lemaitre--Robertson--Walker metric in the natural 3+1 splitting the it yields the expected answer for any admissible surface:
\begin{eqnarray} 
\av{\tilde Z_{ab}} = H(t)\,\delta_{ab}\label{eqflrw},
\end{eqnarray}
i.e. the coarse--grained velocity gradient consist only of the scalar part given by the Hubble parameter. This result is easy to see if we notice that the metric $q_{AB}$ induced on the surface, like any other object in a FLRW solution comoving with the dust, undergoes a homogeneous rescaling in time described by $\dot q_{AB} = 2H(t)\,q_{AB}$. The same metric variation is induced on the image of the embedded surface by the homothetic vector field $v^a = H(t)\,x^a$ in \Ee. It is now straightforward to verify that the restriction of this vector field to $\partial D_t$  gives (\ref{eqflrw}) when plugged into (\ref{eqZab}). 
This result holds for open, closed and flat FLRW metrics alike. 

\section{Time evolution equation for the coarse--grained quantities}
In order to derive the time evolution equations for the coarse--grained velocity gradient, one must
first introduce a preferred time derivative of a tensor field in $\Ee$, which will serve as a generalization of
the covariant derivative along the geodesic in (\ref{eqdotZ}). We define it via 
\begin{eqnarray*}
\Dd T\UD{ab\dots}{cd\dots} &=& \frac{\partial}{\partial t}\,T\UD{ab\dots}{cd\dots} + W\UD{a}{z}\,T\UD{zb\dots}{cd\dots} + W\UD{b}{z}\,T\UD{az\dots}{cd\dots} +\cdots + \nonumber \\  &&- W\UD{z}{c}\,T\UD{ab\dots}{zd\dots} - W\UD{z}{d}\,T\UD{ab\dots}{cz\dots} + \cdots, 
\end{eqnarray*}
where
\begin{eqnarray*}
 W_{ab} = -\frac{1}{V_0}\,\int_{\partial D_t} v_{\leftasym{a}}n_{\rightasym{b}}\,\dd\sigma.
\end{eqnarray*}
$\Dd$ preserves the Euclidean metric in $\Ee$ and has the right narrow tube limit, i.e. it reduces to $\nabla_u$ along the limiting geodesic \cite{korzynski--2009}. 

We now fix a tube of particle worldlines $C_t$, given by functions $\xi^i(\theta^A)$, whose sections $\partial C_t$ are admissible surfaces. We find the isometric embeddings and calculate $\av{\tilde Z_{ab}}$.
This allows for decomposing the \emph{local} $Z_{ij}$, pushed forward to $\partial D_t \subset \Ee$ via $g_p$, and the velocity field $v^a$ into the the coarse--grained, large scale part and local inhomogeneities
\begin{eqnarray}
 g_{p\,*}(Z)_{ab} &=& \av{\tilde Z_{ab}} + \delta Z_{ab} \label{eqdeco1}\\
 v^a &=& \av{\tilde Z\UD{a}{b}}\,x^b + \delta v^a \label{eqdeco2},
\end{eqnarray}
both defined on $\partial D_t$ only.

For the sake of brevity we introduce the following notation for surface integrals of the type of (\ref{eqQsurface}) and (\ref{eqZab}):
\begin{eqnarray*}
 \Nn{X_a}_b = \frac{1}{V_0}\,\int_{\partial D_t} X_a\,n_b\,\dd\sigma.
\end{eqnarray*}
We will also denote by ${\cal R}\left[\,\cdot\,\right]$  the unique inverse of ${\cal P}_\chi$ satisfying
\begin{eqnarray*}
 \Npm{r_{AB}}_{[cd]} = 0
\end{eqnarray*}
for any $r_{AB}$,
plus an irrelevant condition fixing the constant part $E^a$, for example fixing the surface's centroid at the origin.
Now the definition of the coarse--grained $Z_{ij}$ can be written down in a slightly more compact manner
\begin{eqnarray}
 \av{\tilde Z_{(ab)}} = \Npm{\dot q_{AB}}_{ab} = \Npm{2Z_{(ij)}\,\xi\UD{i}{,A}\,\xi\UD{j}{,B}}_{ab} \label{eqcgzab}
 \end{eqnarray}
 in which we have used the time derivative of (\ref{eqqab}) to substitute $2Z_{(ij)}\,\xi\UD{i}{,A}\,\xi\UD{j}{,B}$ for $\dot q_{AB}$.
Note that the combination $\Npm{2\xi\UD{i}{,A}\,\xi\UD{j}{,B}\,\cdot\,}$ plays in our formalism the role of the coarse--graining operator for 
symmetric tensors of rank 2 on $\Sigma_t$.

The evolution equation for $\av{\tilde Z_{ab}}$ takes now the form of
\begin{eqnarray}
 \Dd \av{\tilde Z_{ab}} = -\av{\tilde Z_{ac}}\,\av{\tilde Z\UD{c}{b}} - \av{R_{a0b0}}
 + B_{ab} + \widetilde B_{ab},
\label{eqdQ}
\end{eqnarray}
where 
\begin{eqnarray*}
 \av{R_{a0b0}} =  \Npm{2R_{i0j0}\,\xi\UD{i}{,A}\,\xi\UD{j}{,B}}_{(ab)}
\end{eqnarray*}
is obviously the coarse--grained contraction of the Riemann tensor with $u^\mu\,u^\nu$, considered as a symmetric tensor on $\Sigma_t$ \cite{korzynski--2009}.
We can see that two new terms have appeared in comparison with (\ref{eqdotZ}). The first one is the symmetrized version of the familiar Newtonian backreaction term (\ref{eqnewtonbackr}):
\begin{eqnarray}
 B_{ab} &=& \Nn{\delta v\UD{c}{,c}\,\delta v_{\leftsym{a}}}_{\rightsym{b}} - \Nn{\delta v\UD{c}{\leftsym{,a}}\,\delta v_{\rightsym{b}}}_c \label{eqbackrB}.
\end{eqnarray}
The second one is entirely relativistic and has a more  complicated structure
\begin{eqnarray*}
 \widetilde B_{ab} &=& \Npm{4\av{\tilde Z_{cd}}\,\left(\delta Z\UD{d}{e}\,\chi\UD{e}{\leftsym{,A}} - 
 \delta v\UD{d}{\leftsym{,A}}\right)\,\chi\UD{c}{\rightsym{,B}}}_{ab} + \nonumber \\
 &+&\Npm{2\left(\delta Z_{ce}\,\delta Z\UD{e}{d}\,\chi\UD{c}{,A}\,\chi\UD{d}{,B} - \delta v\UD{c}{,A}\,\delta v\UD{d}{,B}\,
\delta_{cd}\right)}_{ab}.
\end{eqnarray*}
 Just like in the Newtonian case, both backreaction terms are surface integrals divided by volume. In contrast
to (\ref{eqbackrB}) however, $\widetilde B_{ab}$ involves linear terms in perturbations. Both $B_{ab}$ and $\widetilde B_{ab}$ together measure the influence of inhomogeneities on the time evolution of the coarse--grained expansion and shear.

\begin{acknowledgments}
The author would like to thank Lars Andersson, David Wiltshire, Michael Reiris and Gerhard Huisken for useful discussions and comments. The work was supported by the Foundation for Polish Science through the ''Master'' program. 
\end{acknowledgments}


\bibliography{iupaper}


\end{document}